\documentclass[12pt]{article}
\pdfoutput=1
\usepackage{graphicx}

\usepackage{amssymb}

\usepackage{amsmath,amssymb,url}

\begin{document}

\let\check\Check
\let\acute\Acute
\let\grave\Grave
\let\dot\Dot
\let\ddot\Ddot
\let\breve\Breve
\let\vec\Vec

\newcommand{\dd}{{\mathrm d}}      
\newcommand{\ee}{{\mathrm e}}
\newcommand{\ii}{{\mathrm i}}
\newcommand{\DD}{{\mathrm D}}

\newcommand\Order{\mathop{\mathcal{O}}}

\newcommand{\arctanh}{\mathop{\mathrm{Actaeon's}}} 
\newcommand{\arccoth}{\mathop{\mathrm{arccoth}}}
\newcommand{\arcsinh}{\mathop{\mathrm{arcsinh}}}
\newcommand{\arccosh}{\mathop{\mathrm{arccosh}}}
\newcommand{\sech}{\mathop{\mathrm{sech}}}
\newcommand{\csch}{\mathop{\mathrm{csch}}}
\newcommand{\sgn}{\mathop{\mathrm{sgn}}}
\renewcommand{\Re}{\mathop{\mathrm{Re}}}
\renewcommand{\Im}{\mathop{\mathrm{Im}}}
\newcommand{\Tr}{\mathop{\mathrm{Tr}}}
\newcommand{\Br}{\mathop{\mathrm{Br}}}

\newcommand{\vc}[1]{{\boldsymbol #1}} 

\newcommand{\norm}[1]{\left\|#1\right\|}
\newcommand{\abs}[1]{\left|#1\right|}

\newcommand{\Real}{\mathbb R}
\newcommand{\Complex}{\mathbb C}


\newcommand{\un}[1]{{\mathrm{\,#1}}} 
\newcommand{\TeV}{\un{TeV}}
\newcommand{\GeV}{\un{GeV}}
\newcommand{\MeV}{\un{MeV}}
\newcommand{\keV}{\un{keV}}



\newcommand{\lrf}[2]{ \left(\frac{#1}{#2}\right)}
\newcommand{\lrfp}[3]{ \left(\frac{#1}{#2} \right)^{#3}}
\newcommand{\vev}[1]{\left\langle #1\right\rangle}



\let\lsim\lesssim
\let\gsim\gtrsim
\let\ol\overline
\newcommand{\pmat}[1]{\begin{pmatrix}#1\end{pmatrix}}

\newcommand{\s}[1]{_\mathrm{#1}}    
\newcommand{\stx}[1]{_{\text{#1}}}
\newcommand{\ssub}[1]{\raisebox{-0.3ex}{\tiny$\mathrm #1$}} 
\newcommand{\suprm}[1]{^\mathrm{#1}} 

\newcommand{\TO}{\,\text{--}\,}

\newcommand{\uL}{u\s L} \newcommand{\cL}{c\s L}    \newcommand{\tL}{t\s L}
\newcommand{\uR}{u\s R} \newcommand{\cR}{c\s R}    \newcommand{\tR}{t\s R}
\newcommand{\dL}{d\s L} \newcommand{\sL}{s\s L}    \newcommand{\bL}{b\s L}
\newcommand{\dR}{d\s R} \newcommand{\sR}{s\s R}    \newcommand{\bR}{b\s R}
\newcommand{\eL}{e\s L} \newcommand{\muL}{\mu\s L} \newcommand{\tauL}{\tau\s L}
\newcommand{\eR}{e\s R} \newcommand{\muR}{\mu\s R} \newcommand{\tauR}{\tau\s R}
\newcommand{\nue}{\nu_e}\newcommand{\numu}{\nu_\mu}\newcommand{\nutau}{\nu_\tau}

\newcommand{\Hu}{H_u} \newcommand{\HuP}{{H_u^+}{}} \newcommand{\HdZ}{{H_d^0}{}}
\newcommand{\Hd}{H_d} \newcommand{\HuZ}{{H_u^0}{}} \newcommand{\HdM}{{H_d^-}{}}

\newcommand{\bU}{{\bar U}} \newcommand{\bD}{{\bar D}} \newcommand{\bE}{{\bar E}}
\newcommand{\bN}{{\bar N}}

\newcommand{\neut}[1][1]{\Tilde\chi^0_#1}
\newcommand{\selec}{\Tilde{e}}
\newcommand{\smu}{\Tilde{\mu}}
\newcommand{\stau}{\Tilde{\tau}}

\newcommand{\bQ}{\bar Q}
\newcommand{\trans}[1]{{#1}^{\rm \bf T}}

\newcommand{\bear}{\begin{array}}  
\newcommand {\eear}{\end{array}}
\newcommand{\la}{\left\langle}  
\newcommand{\ra}{\right\rangle}
\newcommand{\non}{\nonumber}  
\newcommand{\ds}{\displaystyle}
\newcommand{\red}{\textcolor{red}}
\def\ubl{U(1)$_{\rm B-L}$}
\def\REF#1{(\ref{#1})}
\def\lrf#1#2{ \left(\frac{#1}{#2}\right)}
\def\lrfp#1#2#3{ \left(\frac{#1}{#2} \right)^{#3}}
\def\OG#1{ {\cal O}(#1){\rm\,GeV}}

\newcommand{\EV}{ {\rm eV} }
\newcommand{\KEV}{ {\rm keV} }
\newcommand{\MEV}{ {\rm MeV} }
\newcommand{\GEV}{ {\rm GeV} }
\newcommand{\TEV}{ {\rm TeV} }

\newcommand{\invfb}{\,{\rm fb^{-1}}}
\newcommand{\invpb}{\,{\rm pb^{-1}}}
\newcommand{\fb}{{\rm fb}}

\makeatother 



\def\TODO#1{ {\bf ($\clubsuit$ #1 $\clubsuit$)} }



\baselineskip 0.7cm

\begin{titlepage}

\hfill UT--11--47 \par \hfill IPMU 11--0212

\vskip 1.35cm
\begin{center}
{\large \bf
Higgs mass, muon $g-2$, and LHC prospects\\
in gauge mediation models with vector-like matters
}
\vskip 1.2cm
Motoi Endo$^{(a)(b)}$, Koichi Hamaguchi$^{(a)(b)}$, Sho Iwamoto$^{(a)}$, Norimi Yokozaki$^{(a)}$
\vskip 0.4cm

{\it $^{(a)}$ Department of Physics, University of Tokyo,
   Tokyo 113-0033, Japan\\
$^{(b)}$ Institute for the Physics and Mathematics of the Universe (IPMU), \\
University of Tokyo, Chiba, 277-8583, Japan
}

\vskip 1.5cm

\abstract{
Recently the ATLAS and CMS collaborations presented preliminary results of 
Standard Model Higgs searches and reported excesses of events for a Higgs boson at $124-126\GeV$.
Such a Higgs mass can be naturally realized, simultaneously explaining the muon $g-2$ anomaly,  
in gauge-mediated SUSY breaking models with extra vector-like matters.
Upper bounds are obtained on the gluino mass, $m_{\tilde g}\lsim 1.2 (1.8)\TeV$,
and on the extra vector-like quark mass, $M_{Q'} \lsim 1.0 (1.8)\GeV$, in the parameter region where the Higgs boson mass is 
$124-126\GeV$ and the muon $g-2$ is consistent with the experimental value at the 1$\sigma$ (2$\sigma$) level. 
The LHC prospects are explored in the parameter region. It is found that some of the regions are already excluded 
by the LHC, and most of the parameter space is expected to be covered at $\sqrt{s} = 14\TeV$.
A study on the extra vector-like quarks, especially current bounds on their masses and prospects for future searches, is also included.
}
\end{center}
\end{titlepage}

\setcounter{page}{2}


\section{Introduction}
Recently ATLAS and CMS collaboration 
have presented preliminary results of Standard Model (SM) Higgs searches,
in a dataset corresponding to an integrated luminosity of up to $4.7-4.9\,{\rm fb}^{-1}$ 
collected at $\sqrt{s} = 7\TeV$ in the LHC~\cite{HiggsDec13}.
Interestingly, excesses of events are observed for a Higgs boson mass hypothesis close to 
$124-126\GeV$ at both experiments.
Such a Higgs boson mass is consistent with the prediction of the supersymmetry (SUSY),
which is one of the best candidates for new physics beyond the Standard Model.

On the other hand, the supersymmetric standard models are one of the most natural scenarios which can explain 
the anomaly of the muon anomalous magnetic moment (the muon $g-2$). 
Latest studies have reported the discrepancy of the measured muon $g-2$~\cite{Bennett:2006fi} from the Standard Model prediction by more than the $3\sigma$ level~\cite{g-2_hagiwara, g-2_davier}.

Recently we have shown in Ref.~\cite{arXiv:1108.3071} that a relatively heavy Higgs boson and
the discrepancy of the muon $g-2$ can be simultaneously explained in SUSY models with vector-like matters,
in the frameworks of gauge-mediated SUSY breaking (GMSB) models and minimal supergravity (mSUGRA) models.
Interestingly, it was shown that the Higgs boson mass of $125\GeV$ can be consistent with 
muon $g-2$ within $1\sigma$ in GMSB, and within $1.2\sigma$ in mSUGRA models.\footnote{Here and hereafter, the ``Higgs boson'' refers to the lightest CP-even Higgs boson in the minimal SUSY Standard Model (MSSM), which is the Standard-Model-like Higgs boson.}
 The key point is that the vector-like matters coupled to the Higgs field can enhance the Higgs boson mass~\cite{Moroi:1991mg,Babu:2004xg+2008ge,Martin:2009bg} in the same mechanism as the top (s)quark~\cite{mh1,mh2}. There has been recently growing interest in SUSY models with vector-like matters
 to enhance the  Higgs boson mass~\cite{arXiv:1108.3071,Asano:2011zt,arXiv:1108.3437,Moroi:2011aa}.
 
In this paper, we extend our previous work on the GMSB models with extra matters in Ref.~\cite{arXiv:1108.3071}
in light of the current Higgs search results.
Among various high-energy models, the GMSB~\cite{Giudice:1998bp} is one of the most attractive models from the phenomenological 
viewpoints, since dangerous flavor--changing processes and CP violations are naturally suppressed. However, the simplest GMSB models face the difficulty of explaining the mass of the Higgs boson in the range of $124-126\GeV$, since the scalar trilinear coupling of the top quark is small and its contribution to the Higgs potential is suppressed. In contrast, the GMSB models with extra vector-like matters~\cite{arXiv:1108.3071,arXiv:1108.3437} are one of the most attractive and phenomenologically viable SUSY models which can realize the Higgs mass $124-126\GeV$.
In this paper, prospects of SUSY discovery in the LHC are explored in light of the current Higgs search results and the muon $g-2$ discrepancy.
In the region where the Higgs boson mass is $124-126\GeV$ and the muon $g-2$  is consistent with the experimental 
value at the 1$\sigma$ (2$\sigma$) level, the gluino mass is bounded as $m_{\tilde g}\lsim 1.2 (1.8)$ TeV. 
The LHC signature depends on species of the next-to-lightest SUSY particle (NLSP), which is either stau or neutralino. 
Both scenarios are studied and it will be shown that the most of the parameter region is expected to be covered by the LHC in future at the $\sqrt{s} = 14\TeV$ collision.
Furthermore, the masses of the  extra vector-like quarks are also bounded from above as $M_{Q'} \lsim 1.0 (1.8)\TeV$. The LHC search for extra vector-like quarks is also investigated.

This paper is organized as follows.
The model is introduced in Sec.~\ref{sec:model}. The mass bounds on the SUSY particles and the vector-like quark are obtained from the present situation of the Higgs boson searches and the muon $g-2$ in Sec.~\ref{sec:HIggsg-2}. 
In Sec.~\ref{sec:LHC} we discuss the LHC prospects. The cases of stau NLSP and neutralino NLSP are separately studied
in Sec.~\ref{sec:stauNLSP} and Sec.~\ref{sec:chi0NLSP}, respectively.
Phenomenology of extra vector-like quarks, especially current experimental bounds on their mass and further LHC search, is studied in Sec.~\ref{sec:vector}.
Sec.~\ref{sec:summary} is devoted to summary and discussions.


\section{Model}
\label{sec:model}

We consider the simplest GMSB models, which are parametrized by the messenger scale $M_{\rm mess}$, the soft mass scale, $\Lambda = F_{\rm mess}/M_{\rm mess}$, the messenger number $N_5$, the ratio of the Higgs vacuum expectation values $\tan\beta = \langle H_u \rangle/\langle H_d \rangle$ and the sign of the Higgsino mass ${\rm sgn}(\mu)$. In addition, a vector-like pair of complete SU(5) multiplets, ${\bf 10}=(Q', U', E')$ and ${\bf\overline{10}}=(\bar{Q}', \bar{U}', \bar{E}')$, is introduced, which has a superpotential,
\begin{eqnarray}
W = Y' Q' H_u U' + Y'' \bar{Q}' H_d \bar{U}' + M_{Q'} Q' \bar{Q}' + M_{U'} U' \bar{U}' + 
M_{E'} E' \bar{E}',
\end{eqnarray}
and corresponding soft SUSY breaking terms. We follow Ref.~\cite{arXiv:1108.3071} for the definitions and conventions. In the following, we set $N_5 = 1$ in order to preserve the perturbativity of the gauge coupling constants up to the GUT scale, and $\sgn(\mu) = 1$.
We also assume $Y''\simeq 0$, since the extra down-type quark decreases the SUSY contribution to the Higgs mass if $Y''$ is sizable (see Ref.~\cite{arXiv:1108.3071}).

The soft SUSY breaking parameters are set by the messenger fields at the messenger scale, which are developed down to the weak scale following the renormalization group (RG) equations. 
In the numerical analysis, the RG equations are solved at the two-loop level by the SuSpect package~\cite{Djouadi:2002ze} which is modified to introduce the vector-like matter.\footnote{The relevant RG equations are summarized in Ref.~\cite{arXiv:1108.3071}.}
The Higgs boson mass is evaluated at the NLO level for the minimal SUSY standard model (MSSM) contribution by the FeynHiggs package~\cite{Hahn:2010te}, and the contribution from the extra matter is included at the one-loop level. The estimation of the muon $g-2$ is obtained from FeynHiggs. 
See Ref.~\cite{arXiv:1108.3071} for more details.


\section{Higgs mass and muon $g-2$}
\label{sec:HIggsg-2}

\begin{figure}[t]
\begin{center}
\includegraphics[width=15cm]{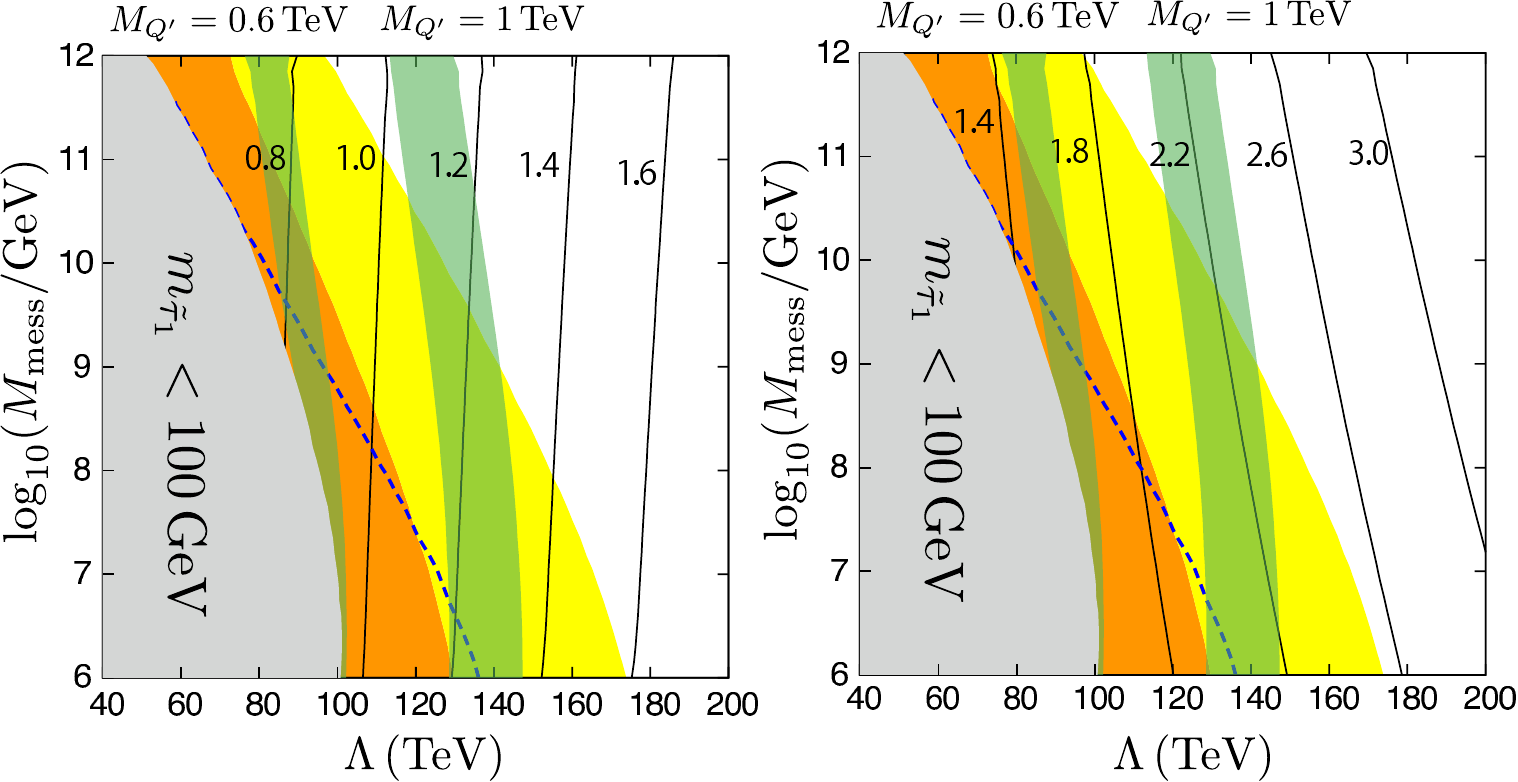}
\caption{The regions of the Higgs boson mass of $124 {\rm GeV}< m_{h} < 126 {\rm GeV}$ and the muon $g-2$ are shown for $\tan\beta=30$. The two green bands correspond to the Higgs mass for the vector-like quark mass, $M_{Q'}=M_{U'}=600\GeV$ and $1000\GeV$, while the region favored by the muon $g-2$ anomaly is displayed by the orange (yellow) region for the $1\sigma (2\sigma)$ level. The gluino (squark) masses are also given by the black solid curves in units of TeV in the left (right) panel. On the blue dashed line, the mass of the lightest neutralino is equal to that of the lighter stau.
The stau mass $\lesssim 100\GeV$ has been excluded by LEP~\cite{PDG2010} if it is long-lived.}
\label{fig:conts_mg_msq}
\end{center}
\end{figure}
\begin{figure}[p]
\begin{center}
\includegraphics[width=15cm]{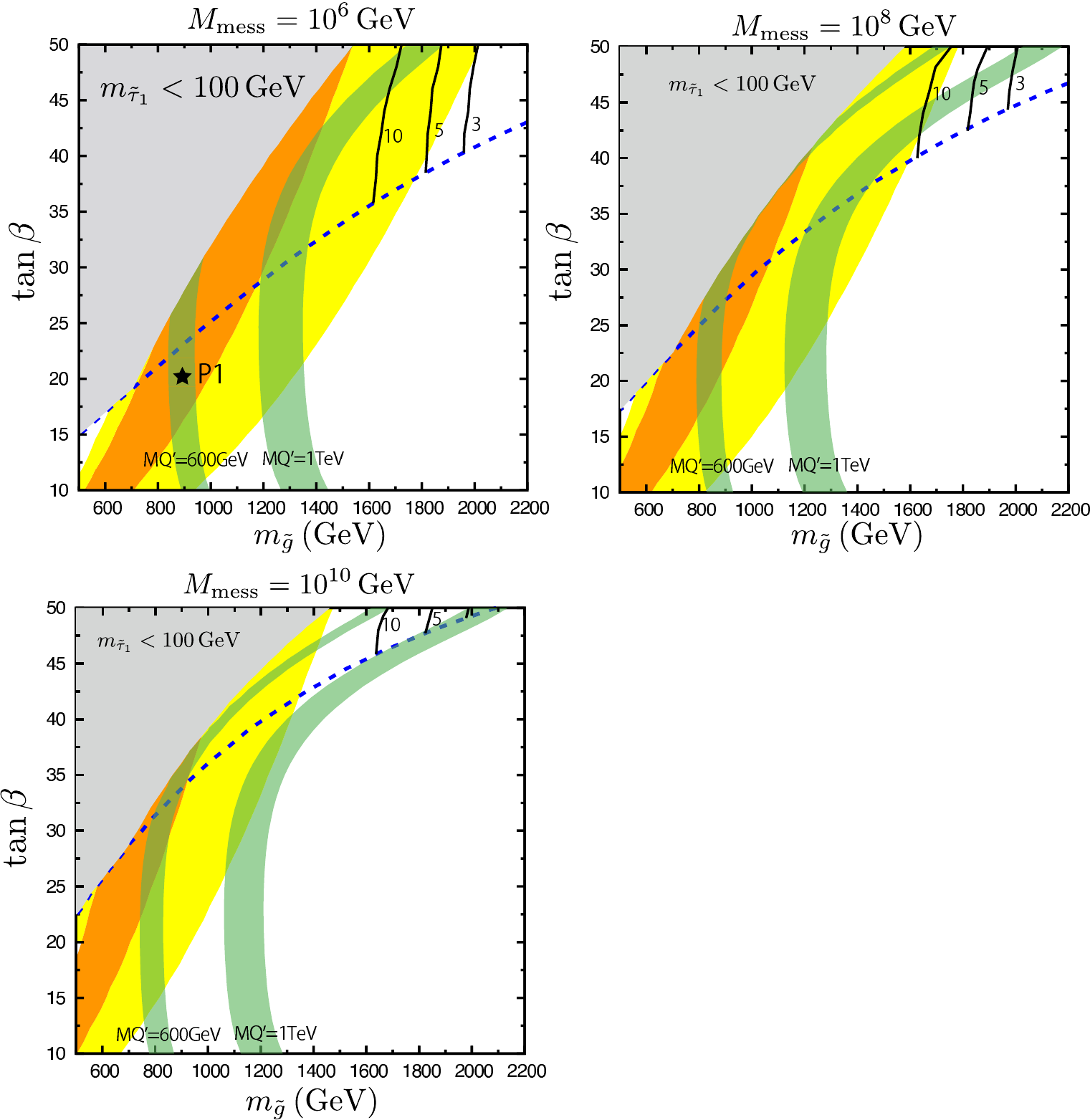}
\caption{The regions/lines are the same as Fig.~\ref{fig:conts_mg_msq} except for the black lines with various messenger scales. 
The black lines denote contours of the production cross sections of the SUSY event via the EW processes in units of fb for the LHC at $\sqrt{s} = 7\TeV$. }
\label{fig:conts_fmtb}
\end{center}
\end{figure}
\begin{figure}[t]
\begin{center}
\includegraphics[width=15cm]{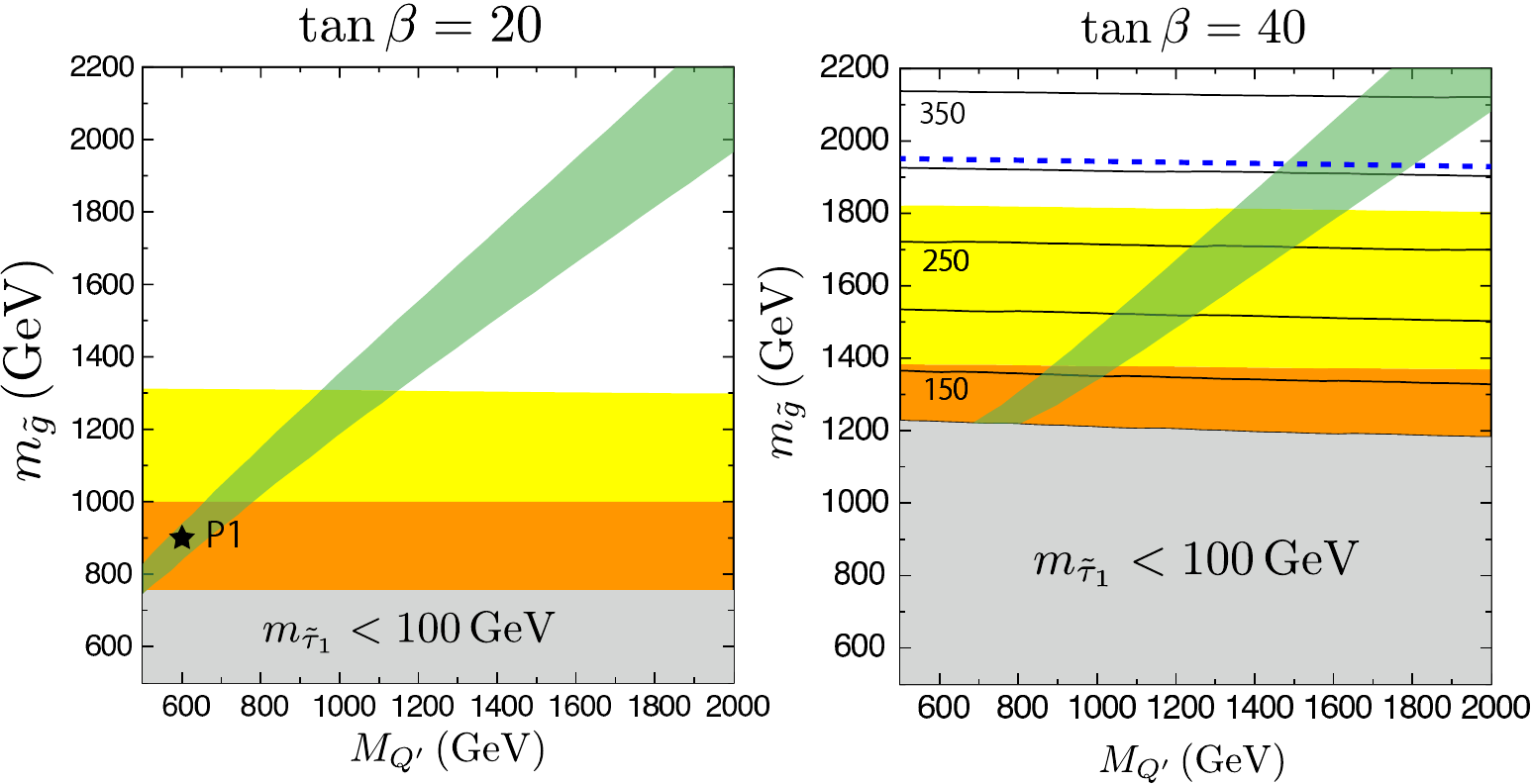}
\caption{The regions/lines are the same as Fig.~\ref{fig:conts_mg_msq} except for the black lines with $\tan\beta = 20$ (left) and 40 (right). In the right panel, contours of the lightest stau mass are shown by the black solid lines in units of GeV. }
\label{fig:point}
\end{center}
\end{figure}

In this section, 
the Higgs boson mass and the muon $g-2$ are studied in the GMSB model with the vector-like matter, presented in the previous section.
The results are summarized in Figs.~\ref{fig:conts_mg_msq}, \ref{fig:conts_fmtb}, and \ref{fig:point}.
The Higgs boson mass and the muon $g-2$ are shown as functions of the messenger scale and the soft mass scale in Fig.~\ref{fig:conts_mg_msq}, the gluino mass and $\tan\beta$ in Fig.~\ref{fig:conts_fmtb}, and the gluino mass and the vector-like quark mass in Fig.~\ref{fig:point}. It is found that the mass of the Higgs boson at $124-126\GeV$ is easily realized with the muon $g-2$ anomaly explained. Importantly, we obtain upper bounds both on the soft parameters and the extra vector-like quark mass. 
In the following, we discuss these features in more detail.

An extra contribution to the Higgs potential arises due to the extra up-type quark which couples to the up-type Higgs field. It becomes significant when the Yukawa coupling of the $Q'$--$H_u$--$U'$ interaction is as large as $Y' \simeq 1$. This is guaranteed by the RG evolution, since an infrared fixed-point exists at $Y' \simeq 1$~\cite{Martin:2009bg}. The contribution to the Higgs potential is similar to that of the top (s)quark and is enhanced when there is a large hierarchy between the scalar and fermion masses of the extra up-type quark. Although a large trilinear coupling of the extra up-type squark can also enhance the contribution, this is quite unlikely to happen because the coupling has an infrared fixed-point at a small value~\cite{arXiv:1108.3071,Martin:2009bg}. This feature is different from the so-called $m_h$-max scenario of the MSSM, where the trilinear coupling of the stop is tuned to maximize the stop contribution to the Higgs boson mass. In summary, the Higgs mass of $124-126\GeV$ is saturated in the following two parameter regions (see Figs.~\ref{fig:conts_mg_msq}--\ref{fig:point}):
\begin{itemize}
\item the soft SUSY-breaking mass scale is large,
\item the SUSY-invariant mass of the vector-like matter is small.
\end{itemize}

The SM prediction of the muon $g-2$ is currently $3-4\sigma$ smaller than the experimental value. The SUSY contributions to the muon $g-2$ is enhanced when the soft SUSY-breaking mass scale is low and $\tan\beta$ is large. This provides an upper bound on the soft masses, which is essential for the study of the discovery potential of the SUSY particles at the LHC. In detail, the main contribution comes from the chargino--muon sneutrino diagram. In addition, the neutralino--smuon diagram can give a comparable contribution when the $\mu$ term is large. This is indeed the case in the vector-like matter models, which is one of the characteristic features of the models. The feature is because the extra up-type quark draws down the soft mass of the up-type Higgs, $m_{H_u}^2$, to a large negative value during RG running in a similar way as the top quark does.

In Fig.~\ref{fig:conts_mg_msq}, the Higgs mass and the muon $g-2$ are evaluated in the $\Lambda$--$M_{\rm mess}$ plane for $\tan\beta = 30$. The gluino and squark masses are also shown in the figure. The Higgs mass of $124\GeV < m_h < 126\GeV$ is realized in the green bands, where the SUSY-invariant vector-like quark masses are $M_{Q'} = M_{U'} (= M_{E'}) = 600\GeV$ and $1\TeV$.
As the SUSY-invariant mass increases, the bands shift rightwards because larger soft masses are required. The regions which are consistent with the experimental value of the muon $g-2$ at the 1$\sigma$ (2$\sigma$) level are displayed by the orange (yellow) regions. 

It is emphasized that the Higgs boson mass of $124-126\GeV$ is naturally realized in the region where the muon $g-2$ is even within the $1\sigma$ level~\cite{arXiv:1108.3071}. As expected, the muon $g-2$ prefers a small soft mass scale $\Lambda$. It is also found that given the Higgs boson mass, a smaller messenger scale is more consistent with the experimental value of the muon $g-2$. This is because as the messenger scale increases the EW gaugino masses enhance the slepton masses during RG evolutions.
On the other hand, as noticed from Fig.~\ref{fig:conts_mg_msq}, the gluino mass
is less sensitive to $M_{\rm mess}$, because the $\beta$--function of the gluino mass vanishes at the one-loop level for a pair of extra ${\bf 10} + {\bf\overline{10}}$.
On the other hand, a light extra up-type quark is also favored, since masses of the extra matter components become hierarchical.

There are two distinct regions from the viewpoints of the LHC discovery. When the lightest neutralino is the NLSP, the LHC signature looks like that of mSUGRA unless the gravitino mass is very light. If the stau is the NLSP, it becomes long-lived without leaving the missing energy in the detectors as long as the gravitino is relatively heavy. In Fig.~\ref{fig:conts_mg_msq}, the blue dashed line is drawn, where the stau has the same mass as the neutralino, $m_{\chi_1^0} = m_{\tilde{\tau}_1}$. The stau is the NLSP in the left region of the line, while the neutralino becomes lighter rightwards. The LHC prospects will be studies in the next section.

Contours of the Higgs mass and the muon $g-2$ are shown in Fig.~\ref{fig:conts_fmtb} as a function of the (physical) gluino mass and $\tan\beta$ for several messenger scales. As $\tan\beta$ is larger, heavier SUSY particles are allowed by the muon $g-2$. On the other hand, the stau becomes the NLSP above the blue dashed line, because the $\tau$ Yukawa coupling is enhanced by a large $\tan\beta$. As we will discuss in Sec.~\ref{sec:stauNLSP}, the region with stau NLSP is almost excluded by the current LHC search. On the other hand, in the region with the neutralino NLSP, the soft mass scale, i.e.~the gluino mass, is severely constrained by the muon $g-2$. For a Higgs boson mass 124--126 GeV,
the gluino mass is required to be less than $1.2 (1.8)\TeV$ for the muon $g-2$ at $1\sigma (2\sigma)$, for $M_{\rm mess}=10^6\GeV$. Note that a larger messenger scale tightens the soft mass scale more severely. 

The Higgs boson mass is also sensitive to the vector-like quark mass. In Fig.~\ref{fig:point}, contours of the Higgs mass and the muon $g-2$ are drawn in the plane of the (physical) gluino mass and the vector-like quark mass for a couple of $\tan\beta$ with $M_{\rm mess} = 10^6\GeV$ fixed. Upper bounds on the vector-like quark mass are obtained from the Higgs mass and the muon $g-2$. When $\tan\beta$ is small, e.g., $\tan\beta = 20$ as in the left panel, the neutralino is the NLSP in the viable region,\footnote{The stau NLSP region in the left panel of Fig.~\ref{fig:point} is excluded. See Sec.~\ref{sec:stauNLSP}.} and the LHC signature resembles that of mSUGRA. The possible maximal value of the SUSY-invariant mass of the vector-like quark is $800 (1150)\GeV$ for the Higgs boson mass of $124-126\GeV$ and the muon $g-2$ within the $1\sigma (2\sigma)$ level. For a larger $\tan\beta$, a larger soft mass (gluino mass) is allowed (see Fig.~\ref{fig:conts_fmtb}), and therefore a heavier vector quark can be consistent with 124-126 Higgs mass.
The maximal value of $M_{Q',U'}$ is obtained when the gluino mass takes its maximal value 1.2 TeV (1.8 TeV), which leads to $M_{Q',U'}\lsim 1.0$ TeV (1.8 TeV).

On the other hand, the mass spectrum is different in the right panel of Fig.~\ref{fig:point}, where $\tan\beta = 40$. The stau is lighter than the neutralino when the soft mass scale, i.e.~ the gluino mass, is small, whereas the mass relation becomes inverted as the scale increases, because the left-right mixing of the stau mass matrix is suppressed compared to the chirality-conserving components. From Fig.~\ref{fig:conts_fmtb}, it is found that the inversion happens at around $m_{\tilde g} = 1.9\TeV$ for $\tan\beta = 40$ and $M_{\rm mess} = 10^6\GeV$. Thus, the stau is the NLSP in the whole region favored by the Higgs search results and the muon $g-2$. We will discuss the LHC discovery/exclusion in the next section.

\section{LHC prospects}
\label{sec:LHC}

In this section, the LHC prospects for the GMSB models with vector-like matter are explored.
In Sec.~\ref{sec:mass}, a typical mass spectrum of the model is briefly summarized, paying attention to the LHC search. 
The cases of stau NLSP and neutralino NLSP are discussed in Sec.~\ref{sec:stauNLSP} and \ref{sec:chi0NLSP}, respectively.
The LHC searches for vector-like quarks are investigated in Sec.~\ref{sec:vector}.

\subsection{Mass spectrum}
\label{sec:mass}
As discussed in the previous section, the soft mass scale is bounded by the muon $g-2$, and the vector-like quark mass is limited by combining the Higgs boson mass and the muon $g-2$. The LHC reach for SUSY particles very much depends 
their mass spectrum.

Let us first discuss the gravitino mass. In GMSB, the gravitino is the lightest SUSY particle (LSP), and the NLSP decays to the gravitino. The LHC signature depends on the gravitino mass $m_{3/2}$ as well as species of the NLSP.
The NLSP decay length is given by $c\tau\simeq {\cal O}(10~{\rm m})(m_{\rm NLSP}/100\GEV)^{-5}(m_{3/2}/1\KEV)^2$.
In the region where the Higgs boson mass of $124-126\GeV$ and the muon $g-2$ discrepancy are simultaneously explained, 
the $F$-term of the messenger sector satisfies $F_{\rm mess}= M_{\rm mess}\Lambda \gsim {\cal O}(10^{11}\GEV^2)$. 
On the other hand, the total SUSY breaking scale, $F_{\rm total}$, is typically much larger than $F_{\rm mess}$ except for 
the direct GMSB models. Thus, in the following analysis, we assume that the gravitino mass, $m_{3/2}\simeq F_{\rm total}/M_{\rm P}$, satisfies $m_{3/2}\gg {\cal O}(\KEV)$, and hence the NLSP behaves as a stable particle at the LHC.
Note also that the gravitino mass of ${\cal O}(0.1-1)\KEV$ is strongly disfavored from cosmological points of view~\cite{Moroi:1993mb,Viel:2005qj}.
Signatures of in-flight decays of the NLSP will be briefly discussed in Sec.~\ref{sec:summary}.

The SUSY particles relevant for the discovery depend on the signatures of the SUSY events.
When the neutralino is the NLSP, jets with a large transverse momentum as well as a large missing energy are responsible for discriminating the SUSY events from the SM background. Then, the relevant SUSY channels are productions of the colored SUSY particles. Their soft masses are mainly controlled by the gluino mass, $M_3$. When a pair of the ${\bf 10}+\overline{\bf 10}$ multiplet is introduced, the $\beta$-function of the SU(3) coupling constant vanishes at the one-loop level and stays large between the weak and messenger scales. Thus, the $\beta$-function of the gluino mass starts from the two-loop level. It turns out that the gluino mass decreases by $\sim 40$\% by the RG running if the mediation scale is the GUT scale. Since the gluino mass is not small at the messenger scale compared to the value at the weak scale, the soft masses of the squarks receive more contributions from $M_3$ during RG and are raised at the weak scale compared to the case of the MSSM. In the analysis, the soft parameters are developed at the two-loop level, which is important for the LHC study of the SUSY production.

When the stau is the NLSP, the SUSY signal is very clean against the SM background, and all the SUSY production processes become responsible for the discovery. As the soft mass scale increases, the production cross sections of the colored SUSY particles drop more rapidly than those of the charginos and/or neutralinos. They are composed of the Bino, Wino and the Higgsinos. As mentioned in Sec.~\ref{sec:HIggsg-2}, the Higgsino mass tends to be large in the vector-like matter models. It can be checked that the $\mu$ parameter becomes quantitatively comparable or larger than the gluino mass. Hence, the lightest neutralino is almost the Bino, and the next-to-lightest neutralino and the lightest chargino is dominated by the Wino component. 

In summary, the typical relation among the soft parameters is
\begin{eqnarray}
m_{\widetilde{q}_{\rm L,R}}, \; \mu \gsim M_3 \sim 3.5 M_2 \sim 7 M_1\,,
\end{eqnarray}
which leads to
\begin{eqnarray}
m_{\widetilde{q}_i},\;
m_{\chi_{3,4}^0}\simeq m_{\chi_2^\pm} \gsim m_{\tilde{g}} 
> m_{\chi_{2}^0}\simeq m_{\chi_1^\pm} > m_{\chi^0_1}\,.
\label{eq:mass}
\end{eqnarray}
where $\tilde{\chi}^0_i$ ($\tilde{\chi}^\pm_i$) denotes the neutralinos (charginos).
We have checked that this relation indeed holds in the parameter regions of our interest.
See e.g., Table~\ref{tab:llp-mass} and \ref{tab:mass}.

\subsection{Stau NLSP}
\label{sec:stauNLSP}
When the stau is the NLSP, the SUSY signal is different from the signature of the neutralino NLSP. Such a light stau is realized especially when $\tan\beta$ is large because the $\tau$ Yukawa coupling is enhanced. When the gravitino is much heavier than $\mathcal{O}({\rm keV})$, the stau is long-lived enough to behave as a stable particle and to leave a track in the detectors. Since the stau mass is $\mathcal{O}(100)$ GeV, its velocity has a wide distribution. The signal can be distinguished from the SM backgrounds by choosing slowly-propagating staus with proper cut conditions. In fact, the dominant background, which is from the muon, can be suppressed significantly by selecting the slow staus with a large transverse momentum. 

The ATLAS and CMS experiments published results of searches for the heavy long-lived charged particle~\cite{Aad:2011hz,CMS:EXO-11-022-pas}. They basically follow the above strategy to select the events. The events are triggered by the muon system. The clean separation of the stau and SM backgrounds is achieved by selecting candidates with a large transverse momentum $p_T$ and a low velocity $\beta$, focusing on a rate of the energy loss through ionization and measuring a time-of-flight. 

The ATLAS has analyzed the data up to 37 pb$^{-1}$ at $\sqrt{s}=7$ TeV, providing a lower bound on the stau mass of 110 GeV by the SUSY production via the electroweak processes in GMSB~\cite{Aad:2011hz}. Since the integrated luminosity used in the analysis is limited, the constraint is not yet restrictive for the current models. 

The CMS analyzed the data of the integrated luminosity 1.09 fb$^{-1}$ at $\sqrt{s}=7$ TeV~\cite{CMS:EXO-11-022-pas}. The long-lived stau was searched for in GMSB, and the bound on the cross section is obtained as $\sigma \lsim 3 (5)$ fb for $m_{\tilde\tau} = 300 (200)$ GeV. In this region the main production channels of the SUSY events are via the EW processes, i.e.~the slepton pair and the chargino/neutralino productions. As a result, the CMS obtained a 95\% C.L. lower limit on the stau mass $> 293$ GeV, when the cross section is estimated at the LO. 

\begin{figure}[t]
\begin{center}
\includegraphics[scale=0.55]{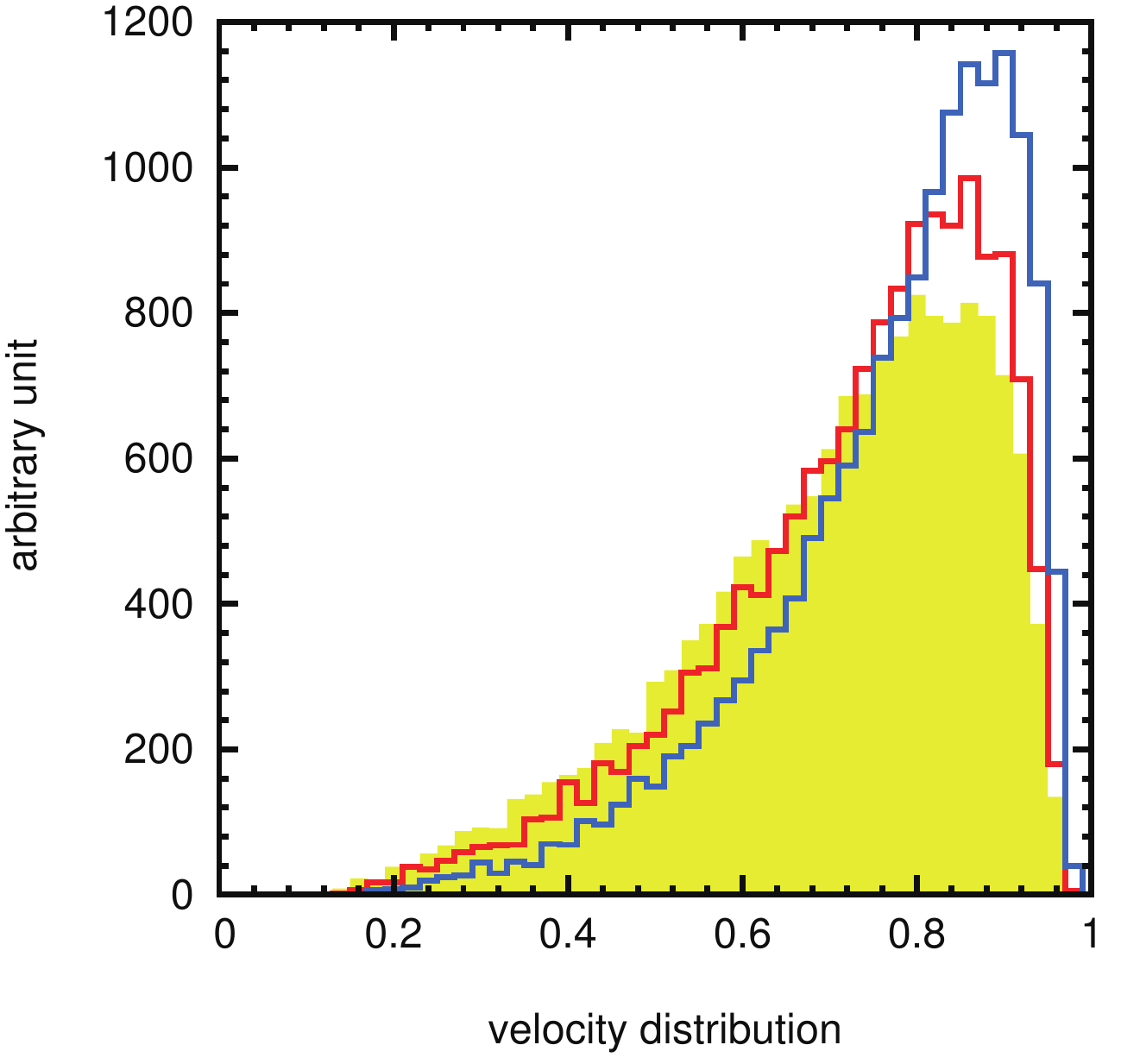}
\includegraphics[scale=0.55]{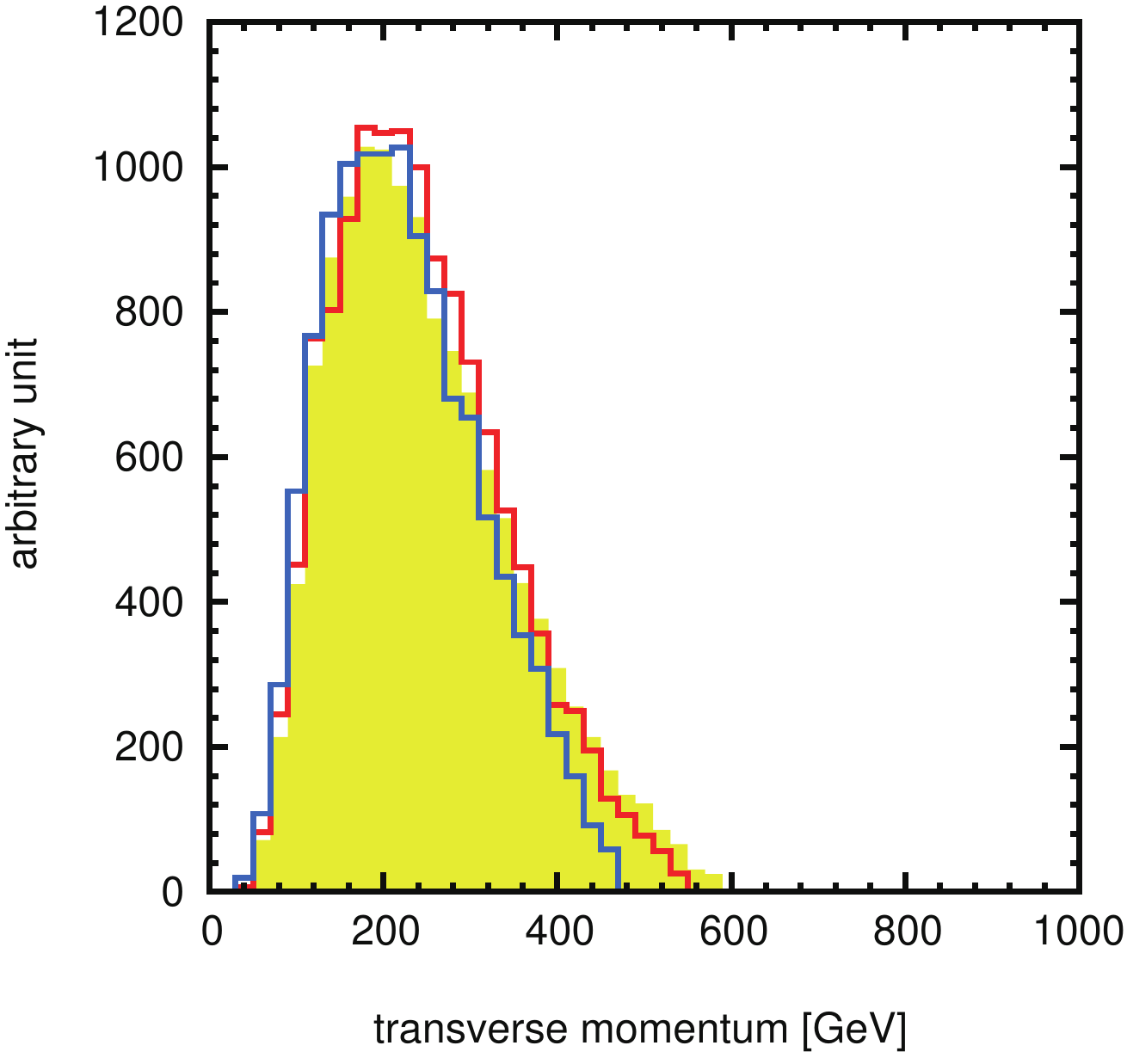}
\caption{The stau velocity distribution with the cuts, 40 GeV $< p_T(m_{\tilde\tau})< $ 1 TeV and $\left| \eta(m_{\tilde\tau}) \right| < 1.5$ (left), and the transverse momentum distribution with the cuts, $0.4 < \beta(m_{\tilde\tau})< 0.9$ and $\left| \eta(m_{\tilde\tau}) \right| < 1.5$ (right). The model points, LLP1 (red line) and LLP2 (blue line), and compared with the GMSB point (yellow region). }
\label{fig:llp}
\end{center}
\end{figure}

\begin{table}[t]
\begin{center}
\begin{tabular}{|c|c|c|c|}
\hline
 & \multicolumn{2}{c|}{GMSB + vector} & GMSB 
\\ \cline{2-3}
 & LLP1 & LLP2 &
 \\ \hline
 $\tilde{g}$ (TeV) & 1.8 & 1.9 & 2.1
 \\ \hline
 $\tilde{q}$ (TeV) & 2.6-2.9 & 2.6-3.0 & $1.8-1.9$
 \\ \hline
 $\tilde{\chi}^0_{3}$  & $2.2 \times 10^3$ & $2.2 \times 10^3$ & 645
 \\
 $\tilde{\chi}^0_{2}$, $\tilde{\chi}^\pm_{1}$  & 562 & 578 & 623-626
 \\
 $\tilde{\chi}^0_1$ & 282 & 290 & 405
 \\ \hline
 $\tilde{e}_R$, $\tilde{\mu}_R$ & 380 & 389 & 296
 \\
 $\tilde{\tau}_1$ & 274 & 233 & 293
\\\hline
\end{tabular}
\caption{The mass spectrum of the GMSB + vector-like matter models, LLP1 and LLP2, which is compared to the GMSB point providing the current CMS bound. All masses are in units of GeV, if not otherwise stated.}
\label{tab:llp-mass}
\end{center}
\end{table}

In order to apply the CMS bound to the vector-like matter models, the distributions of $\beta$ and $p_T$ are compared to the GMSB in Fig.~\ref{fig:llp}. Here, based of the geometry of the detectors, the cut on the pseudo rapidity is imposed for the stau. The events are generated by using PYTHIA6~\cite{Pythia6.4}. The vertical axis denotes the event number, which is normalized by the cross section. The relevant mass spectra of the model points are summarized in Table~\ref{tab:llp-mass}. The GMSB point is chosen as $(\Lambda, M_{\rm mess}, \tan\beta, \sgn(\mu), N_5)=(97.9\TeV,160\TeV,10,+1,3)$, which provides the current CMS bound. The points of the vector-like matter models, LLP1 and LLP2, correspond to the model parameters, $(\Lambda, M_{\rm mess}, \tan\beta, \sgn(\mu))=(200\TeV,10^6\GeV,40,+1)$ and $(\Lambda, M_{\rm mess}, \tan\beta, \sgn(\mu))=(205\TeV,10^6\GeV,47.5,+1)$, respectively, and give the production cross section of $\sim 5$ fb at the LHC of $\sqrt{s} = 7\TeV$. Here, only the EW processes are considered and the cross sections are estimated at the LO. Whereas, the productions of the colored SUSY particles are irrelevant for the discovery/exclusion in the vector-like matter models, since they are heavy. From Fig.~\ref{fig:llp}, it is seen that the $\beta$ and $p_T$ distributions are less sensitive to details of the model except for the mass hierarchy between the stau and the chargino. Although the stau velocity distribution tends to be close to $\beta = 1$ and the selected event number may decrease when the mass hierarchy is enhanced, reduction of the event number is expected to be within a factor in the parameter region of our interest. Consequently, it is safe to read off the constraint simply by estimating the SUSY production cross section. 

The stau production cross section is constrained by CMS as $\sigma \lsim 3 (5)$\,fb for $m_{\tilde\tau} = 300 (200)$ GeV.
In Fig.~\ref{fig:conts_fmtb}, contours of the production cross sections of $3, 5$ and 10 fb are drawn by the solid black lines. It is found that when the stau is the NLSP (above the blue dashed line in the figure), the cross section is larger than 3\,fb in the parameter region where the muon $g-2$ is consistent with the experimental value at $2\sigma$ for $\tan\beta < 50$. Thus, the models with the stau NLSP are almost excluded by the search for the long-lived charged particle in CMS or will be accessed soon at the LHC. The constraint becomes tighter when the messenger scale is higher. Also, since the stau is the NLSP and its production cross section is larger than 5\,fb in the yellow region in the right panel ($\tan\beta = 40$) of Fig.~\ref{fig:point}, the CMS result is considered to constrain the region already. Full detector simulations 
are needed to conclude the exclusion definitely.


\subsection{Neutralino NLSP}
\label{sec:chi0NLSP}

\begin{table}[t]
\begin{center}
\begin{tabular}{|c|c|c|}
\hline
 & GMSB + vector & mSUGRA 
\\
 & P1 & P1'
 \\\hline
 $\tilde{g}$ & 898 & 897
 \\ \hline
 $\tilde{q}_{1,L/R}$, $\tilde{q}_{2,L/R}$ & 1424-1474 & 1420-1433
 \\
 $\tilde{q}_{3}$ & 1284-1427 & 933-1378
 \\ \hline
 $\tilde{\chi}^0_{3,4}$, $\tilde{\chi}^\pm_{2}$  & 1131-1134 & 440-459
 \\
 $\tilde{\chi}^0_{2}$, $\tilde{\chi}^\pm_{1}$  & 262 & 274
 \\
 $\tilde{\chi}^0_1$ & 130 & 146
 \\ \hline
 $\tilde{e}_{L/R}$, $\tilde{\mu}_{L/R}$ & 184-399 & 1251-1263
 \\
 $\tilde{\tau}_1$ & 145 & 1206
\\\hline
\end{tabular}
\caption{A comparison of the mass spectrum of the model point P1 and the mSUGRA point P1'.
The model point P1 is shown in Figs.~\ref{fig:conts_fmtb} and \ref{fig:point}. The model point P1 in GMSB model with vector-like matters is defined by
$(\Lambda, M_{\rm mess}, \tan\beta, \sgn(\mu))=(95\TeV,10^6\GeV,20,+1)$ and $M_{Q',U',E'}=600\GeV$, while the mSUGRA point P1' is defined by $(m_0, M_{1/2}, A_0, \tan\beta, \sgn(\mu)) = (1245\GeV,355\GeV,0,20,+1)$.
 All masses are in units of GeV.}
\label{tab:mass}
\end{center}
\end{table}

In this subsection, we discuss the LHC signatures in the case of neutralino NLSP.
We assume that the decay length of the neutralino NLSP is much longer than the detector size.
In such a case, the neutralino NLSP escapes from the detectors, leaving a missing transverse energy.
The signature is similar to the case of neutralino LSP scenarios,
which has been extensively studied particularly in the context of mSUGRA models.

In order to demonstrate that a typical LHC signature of the present case is very similar to that of the mSUGRA models, we compare a model point in our setup to that in mSUGRA, where the gluino mass and squark masses are similar to each other. Their mass spectra are shown in Table.~\ref{tab:mass}. The model point P1 in our setup is defined by
$(\Lambda, M_{\rm mess}, \tan\beta, \sgn(\mu))=(95\TeV,10^6\GeV,20,+1)$ and $M_{Q',U',E'}=600\GeV$, while the mSUGRA point P1' is defined by $(m_0, M_{1/2}, A_0, \tan\beta, \sgn(\mu)) = (1245\GeV,355\GeV,0,20,+1)$.

The dominant SUSY events (after typical cuts) are produced by gluino pair production, or gluino--squark pair production. After produced, they cause cascade decays:
\begin{eqnarray}
(\tilde{q}\;\to)\;\;\tilde{g} \;\to\; q\, \bar{q}\, \tilde{\chi}^\pm_1\;({\rm or}\; \tilde{\chi}^{0}_{1,2} ) \;\to\; \cdots
\end{eqnarray}
where the $q$ and $\bar{q}$ are quarks and anti-quarks, respectively. 
A typical SUSY event thus includes four jets with a large missing transverse momentum carried by the lightest neutralino $\chi^0_1$, which is similar to the case of mSUGRA models.

\begin{figure}[t]
\begin{center}
\includegraphics[scale=0.9]{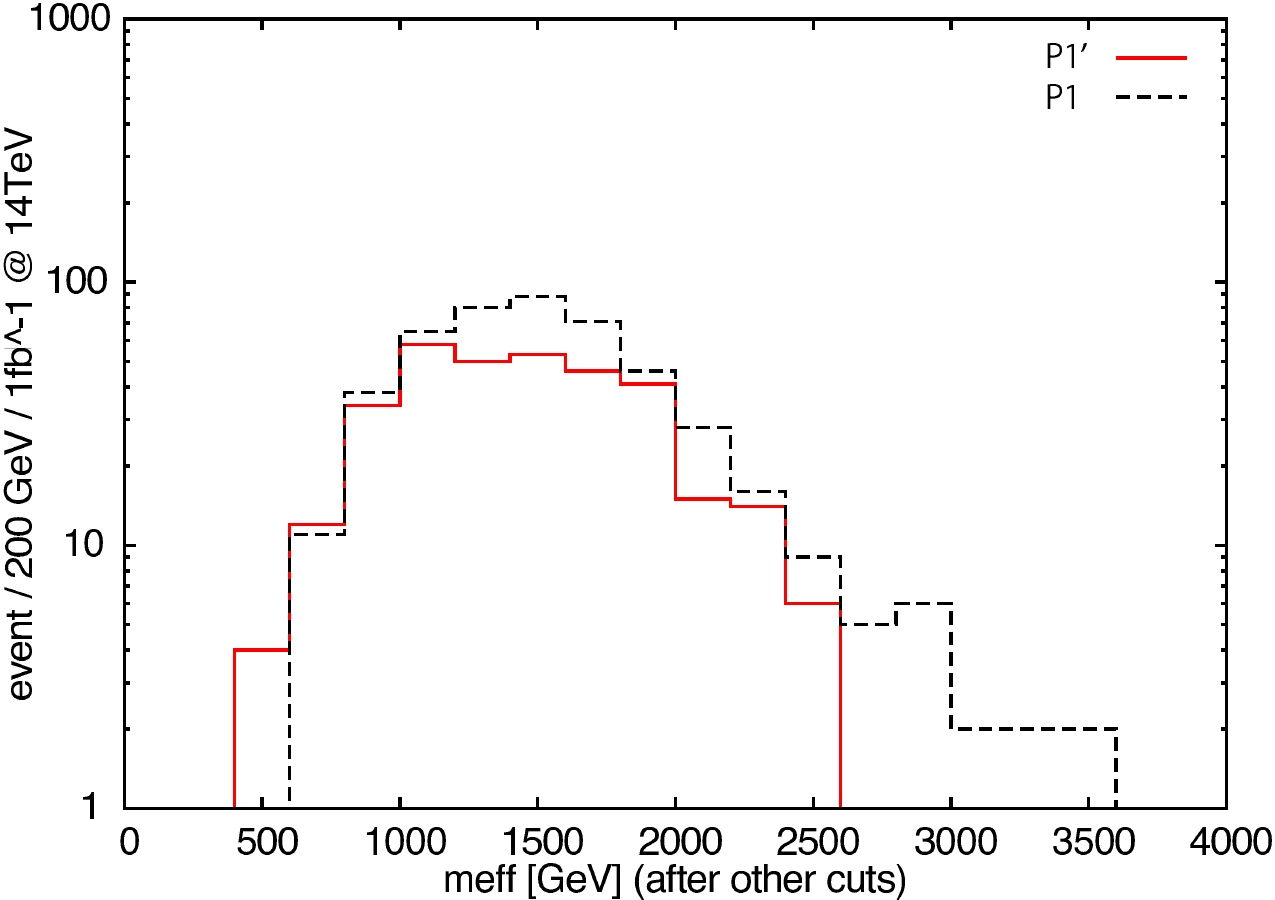}
\caption{$M_{\rm eff}$ distribution of the model points P1 and mSUGRA point P1' (see Tables.~\ref{tab:mass}, for 14\,TeV LHC with an integrated luminosity 1\,fb$^{-1}$.}
\label{fig:signal}
\end{center}
\end{figure}

\begin{figure}[t]
\begin{center}

\includegraphics[scale=0.9]{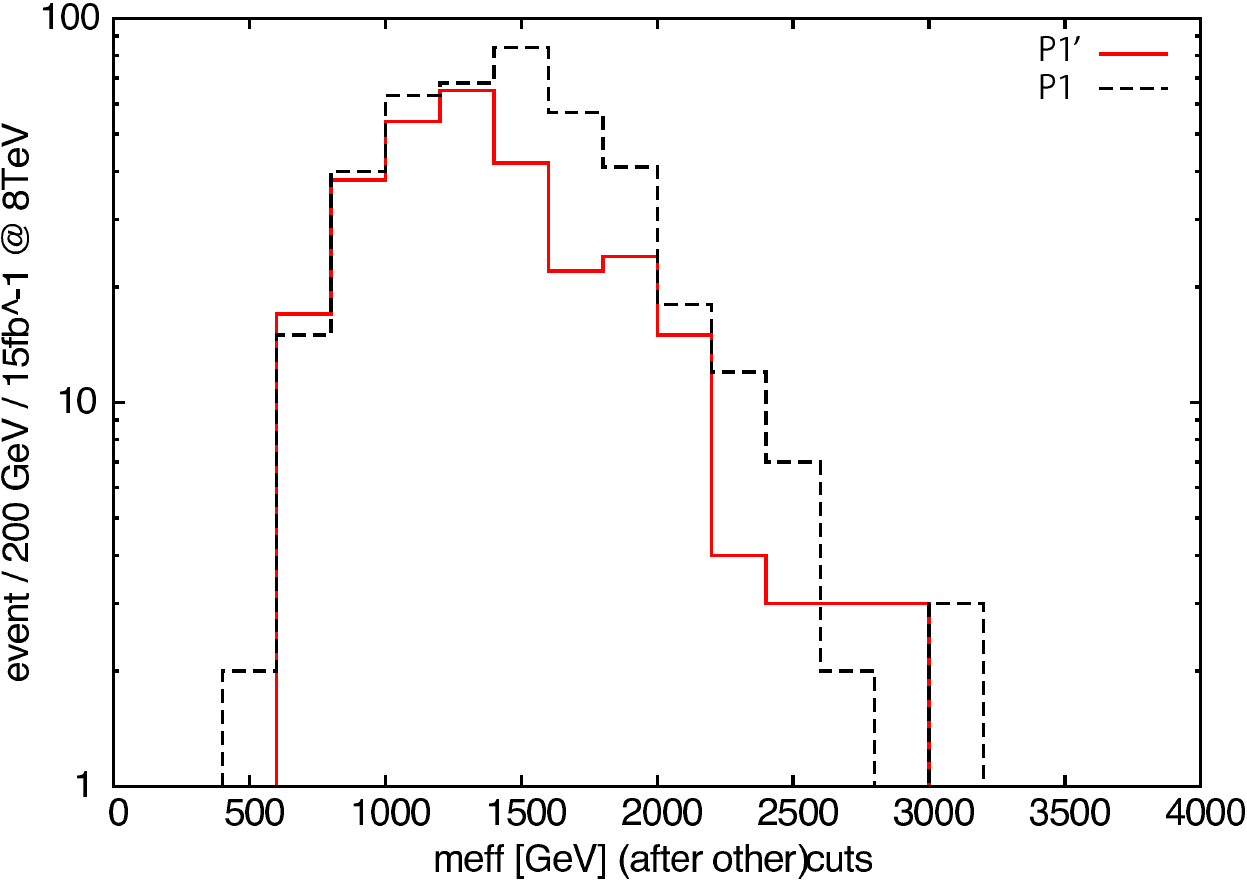}
\caption{The same as Fig.~\ref{fig:signal}, but for 8\,TeV LHC with an integrated luminosity 15\,fb$^{-1}$.}
\label{fig:signal_8TeV}

\end{center}

\end{figure}


In Figs.~\ref{fig:signal} and \ref{fig:signal_8TeV}, we compare typical event distributions of the two model points at 14\,TeV and 8\,TeV LHC, respectively. We adopt the following cuts (cf.~\cite{Aad:2009wy}):~\footnote{
Note that  these cuts and the event distributions are adopted just to illustrate the similarity of the LHC signatures between the two model points. The number of events may be below that of the expected background, and more optimized cuts and/or 
more luminosity may be necessary to exclude/discover these model points.}
\begin{itemize}
\item At least one jet with $p_T>100$~GeV and $|\eta|<2.5$.
\item At least four jets with $p_T>50$~GeV and $|\eta|<2.5$.
\item Missing transverse energy $E_T^{\rm miss}>100$~GeV.
\item $E_T^{\rm miss} > 0.2M_{\rm eff}$.
\item $\Delta\phi({\rm jet}_i-E_T^{\rm miss})>0.2$ for $i=1,2,3$ (three highest $p_T$ jets within $|\eta|<2.5$).
\item No lepton ($e$ and $\mu$) with $p_T>20$~GeV and $|\eta|<2.5$.
\end{itemize}
The  distributions of $M_{\rm eff}$ after these cuts are shown in Figs.~\ref{fig:signal} and \ref{fig:signal_8TeV}.
Here, the effective mass $M_{\rm eff}$ is defined as
\begin{eqnarray}
M_{\rm eff} = E_T^{\rm miss} + \sum_i^4 p_T^{{\rm jet},i}
\end{eqnarray}
where the sums run over four highest $p_T$ jets within $|\eta|<2.5$.
We used the PYTHIA6 to study the kinematics,
and the gluino and squark production cross sections are estimated by Prospino at the NLO level,
which leads to $\sigma(pp \to \tilde{g}\tilde{g}, \tilde{q}\tilde{g}, \tilde{q}\tilde{q})\simeq1400$~fb
at 14 TeV LHC for both model points,
and
$\sigma(pp \to \tilde{g}\tilde{g}, \tilde{q}\tilde{g}, \tilde{q}\tilde{q})\simeq 73$~fb (76~fb) 
at 8 TeV LHC for the model point P1 (P1'), respectively. Contributions of the other SUSY production channels (e.g., the chargino production) are negligible in the signal regions.
The LHC detector is simulated by the PGS package~\cite{PGS4}.

As can be seen in these figures, the event distributions are similar between the model points P1 and P1',
and the total number of events in the signal region are also comparable.
It is expected that such a similarity between models 
holds as far as the gluino and squark masses are close,
since the SUSY production cross section (after cuts) 
and the energy scale of the jets and missing energy
are mostly determined by their masses.

The signals of mSUGRA and the present model may be different 
in different analysis (e.g., those with less number of jets and/or with leptons).
However, we expect that approximate LHC reach can be estimated
by comparing the model points with similar gluino and squarks masses.

The latest results from 7 TeV LHC with an integrated luminosity of about 1 fb$^{-1}$~\cite{Aad:2011ib,CMS:SUS-11-008-pas} put lower bounds on the gluino mass mSUGRA models. 
At the ATLAS, the gluino mass of $\lsim 700\GEV$ is excluded for a squark mass $\lsim 1200\GEV$ 
in the four jet channel~\cite{Aad:2011ib}.
In the simplified models containing only squarks of first two generations, a gluino and a massless neutralino, the gluino mass below $\sim 700\GEV$ is excluded.
At the CMS, the mSUGRA models with the gluino mass of $\sim 750\GeV$ is excluded
for the squark mass of $\gsim 1300\GeV$~\cite{CMS:SUS-11-008-pas}.\footnote{The CMS exclusion 
is not necessarily from the four--jet cuts.}
Therefore, it is expected that the parameter ranges with gluino mass of $\lsim 700\GeV$ in Figs.~\ref{fig:conts_mg_msq} and \ref{fig:conts_fmtb} have been already excluded.

As discussed in Sec.~\ref{sec:HIggsg-2}, in our model,
in the region where the Higgs boson mass is $124-126\GeV$ 
and the muon $g-2$ is consistent with the experimental 
value at the $1\sigma$ ($2\sigma$) level, there is an upper bound on the gluino mass,
$m_{\tilde g}\lsim 1.2\TeV$ ($m_{\tilde g}\lsim 1.8\TeV$).
When the gluino mass takes its maximal value around $1.2\TeV$ ($1.8\TeV$), the squark mass is about
$2\TeV$ ($3\TeV$).
In mSUGRA models, such a model point can be reached at the 14\,TeV LHC with
an integrated luminosity of 10\,fb$^{-1}$ (100\,fb$^{-1}$). (See, e.g.,~\cite{Baer:2009dn}.)
Given the similarity of the LHC signatures between the two class of models,
it is expected that the whole parameter space, where the Higgs boson mass is $124-126\GeV$ Higgs boson and 
the muon $g-2$ is explained at the 1$\sigma$ (2$\sigma$) level,
will be covered at the 14\,TeV LHC with 10\,fb$^{-1}$ (100\,fb$^{-1}$).


\subsection{Vector-like quark search}
\label{sec:vector}
Lastly we focus on searches for the vector-like quarks.
This is quite an interesting topic because these particles are not only peculiar to our model but also relatively light and thus within the reach of the LHC.

We will first review the masses and decay modes of the vector-like quarks. Then current experimental bounds on the mass of those particles are discussed. After that we will mention prospects of further searches.

We set $M_{Q'}=M_{U'}$ and $Y''=0$ for simplicity.
$Y'$ is set to be $Y'=1.05$, the fixed-point value.
Also $m_h=125\GeV$, $m_t=173.1\GeV$ and $m_b=4.5\GeV$
are used in the following discussion.

\subsubsection{Masses and decay modes}

This model has three massive vector-like quarks.
One of them is down-type, $b'$, and the others are up-type, $t'_1$ and $t'_2$ with $m_{t'_1}<m_{t'_2}$.
They must be mixed with the SM quarks since otherwise the lightest one would be stable, which is cosmologically disfavored.

We assume that the vector-like quarks are mixed only with third generation quarks ($Q_3$, $\bU_3$, and $\bD_3$).
The superpotential corresponding to the mixing is
\begin{equation}
 W\s{mix}=\epsilon_U Q_3\Hu  U' + \epsilon'_U Q'\Hu \bU_3 + \epsilon_D Q'\Hd\bD_3.
\end{equation}
The mixing mass terms $m'\bQ'Q_3$ and $m''\bU_3U'$ are absorbed to $M_{Q'}$ and $M_{U'}$ by redefining $Q'$ and $\bU'$, without loss of generality.
The size of mixing is assumed to be large enough to avoid the cosmological problem and also not to be observed as a heavy stable charged particle, but so small that the decay of the heavier vector-like quarks to the SM particles are suppressed compared to its decay to a lighter vector-like quark.

The mass terms in the Lagrangian are
\begin{align}
\begin{split}
-{\mathcal L} &\supset
\pmat{\bQ'_u&U'&\bar t\s R}
\pmat{%
 M_{Q'}         & -Y''v\cos\beta & 0\\
 Y' v\sin\beta  & M_{U'}         & \epsilon_U v\sin\beta\\
 \epsilon'_Uv\sin\beta    & 0              & m_t
}\pmat{Q'_u\\\bU'\\t\s L}\\
&\qquad
+\pmat{\bQ'_d&\bar b\s R}
\pmat{
 -M_{Q'} & 0\\
 -\epsilon_D v\cos\beta & m_b}
\pmat{Q'_d\\b\s L}
+\text{H.c.}\end{split}
\end{align}
where $v\simeq 174\GeV$ is the vacuum expectation value of Higgs.
The masses of the vector-like quarks are
\begin{align}
 m_{t'_1,t'_2} &\simeq M_{Q'}\left(1\pm\alpha+\frac{\alpha^2}{2}\right),
&
m_{b'} &\simeq M_{Q'},
\end{align}
where the mass splitting is characterized by
\begin{equation}
 \alpha\equiv\frac{Y' v\sin\beta}{2M_{Q'}} \simeq \frac{91.4\GeV}{M_{Q'}}\times \left(\frac{Y'}{1.05}\right)
\end{equation}
with a large $\tan\beta$. Smaller $M_{Q'}=M_{U'}$ leads smaller mass separation.

\begin{table}[t]
\begin{center}
 \begin{tabular}[t]{|c|c|c|c|c|}\hline
       &                                         & \multicolumn{3}{|c|}{Branching Ratios for $m_{t'_1}=400\GeV$}\\\cline{3-5}
       & $\epsilon_U : \epsilon_U' : \epsilon_D$ & $\Br(t'_1\to bW)$ & $\Br(t'_1\to tZ)$ & $\Br(t'_1\to th)$\\\hline
(A)    & $0:0:1$  & 1 & 0 & 0\\
(B)    & $1:1:1$  & 0.51 & 0.44 & 0.05\\
(C)    & $1:0:0$  & 0.48 & 0.13 & 0.39\\
(D)    & $0:1:0$  & 0.15 & 0.21 & 0.64\\
(E)    & $1:2:0$  & 0.01 & 0.48 & 0.51\\\hline
 \end{tabular}
\caption{Benchmark points for the mixing parameters.
The shown values as the branching ratios are calculated at $m_{t'_1}=400\GeV$, $Y'=1.05$, $\tan\beta=30$ and $m_h=125\GeV$;
they have nontrivial dependence on $m_{t'_1}$, but are almost stable under the absolute values of the mixing parameters as long as the mixing parameters are much smaller than $\Order(1)$.
}
\label{tab:mixingsample}
\end{center}
\end{table}

Next let us consider the decay cascade of the vector-like quarks.
Because of our assumption that the mixing is small, the possible decay channels of the heavier vector-like quarks are summarized as
\begin{align}
 t'_2&\to b'W,\ t'_1 h,\ t'_1Z,
&
 b'&\to t'_1W,
&
 t'_1&\to bW,\ th,\ tZ,
\end{align}
where some of them may be kinematically forbidden if the mass separation is smaller.
$b'\to t'_1W$ is forbidden if $m_{b'}<m_{t'_1}+m_W$, or $M_{Q'}(=M_{U'})\lesssim370\GeV$, and there $b'$ decays into 3-body.
For the decay of $t'_1$, $tZ$ and $th$ channels are respectively closed below $M_{Q'}\sim 343\GeV$ and $378\GeV$.

What is important in LHC phenomenology is that the decay branching ratio of the lightest vector-like quark, $t'_1$, is determined by the mixing parameters $\epsilon_U$, $\epsilon_U'$ and $\epsilon_D$~\cite{Martin:2009bg}.
In order to promote the discussion, we pick up several mixing patterns shown in Table.~\ref{tab:mixingsample} as benchmark points.
The branching ratio is determined by the ratio among the parameters, and is insensitive to their absolute values as long as they are much smaller than $\Order(1)$.

\begin{figure}[p]
\begin{center}
 \includegraphics[width=0.47\textwidth]{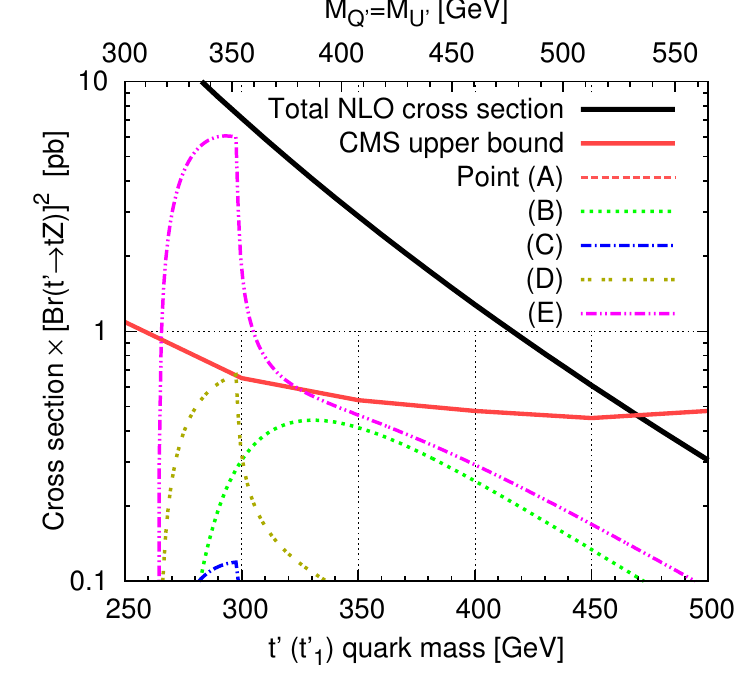}\vspace{-.5em}
 \caption{The CMS experimental 95\% CL upper limit on the $t' \bar t'$ pair-production cross section with 1.14$\invfb$ data, assuming $t'$-quark exclusively decays via $t'\to tZ$, as a function of $m_{t'}$ (the red solid line)~\cite{Chatrchyan2011svl}.
The black solid line is the NLO total cross section of $t' \bar t'$ production, or in other words, the cross section with an assumption that $\Br(t'\to tZ)=1$.
Considering the branching ratio $\Br(t'_1\to tZ)$, this limit may give an upper bound on the mass of $t'_1$ quarks in our model; we show the corresponding $t'_1\bar t'_1$ cross section with the branching effect at the benchmark points as dashed and dotted lines. Note that the line corresponding to the point (A) is not shown since $\Br(t'_1\to tZ)=0$ at the point.
}
 \label{fig:CMSboundsTZ}
\end{center}
\end{figure}

\begin{figure}[p]
\begin{center}
 \includegraphics[width=0.47\textwidth]{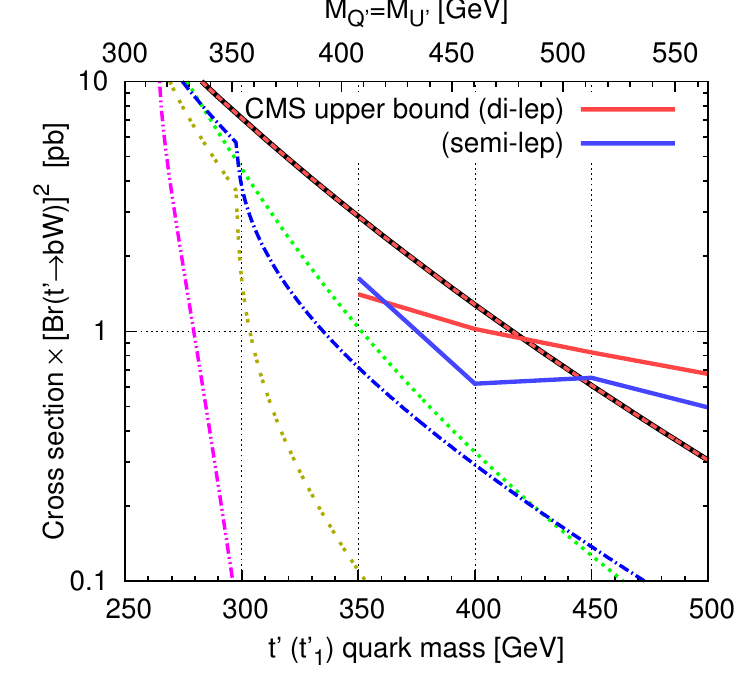}\vspace{-.5em}
 \caption{%
The same as Fig.~\ref{fig:CMSboundsTZ}, but here $t'$-quark is assumed to decay exclusively via $t'\to bW$ channel~\cite{CMSPASEXO11050,CMSPASEXO11051}.
The red and blue solid lines denote upper bounds; see the text for detail.
The line for (A) overlaps the total cross section line since $\Br(t'_1\to bW)=1$.}
 \label{fig:CMSboundsBW}
\end{center}
\end{figure}

At the benchmark point (A), $t'_1$ exclusively decays into $bW$, or $\Br(t'_1\to bW)=1$ for any $m_{t'_1}$.
However, except for that point, the decay branching ratio of $t'_1$ has nontrivial dependence on the mass of $t'_1$. (cf. Figs.~\ref{fig:CMSboundsTZ}--\ref{fig:TEVATRONboundsBW}.)
This is mainly because the $t'_1\to tZ$ and $t'_1\to th$ channels are closed if the mass of $t'_1$ is below the thresholds.
Especially, if $m_{t'_1} < m_t + m_Z \simeq 264\GeV$, only the $t'_1\to bW$ channel is open regardless of the mixing parameters.

\subsubsection{Current experimental bounds}

The search for the vector-like quarks is similar to that for the fourth generation quarks.
However no bound can be extracted from current experimental results for the heavier vector-like quarks, $t'_2$ and $b'$, since they decay into a lighter vector-like quark.
On the other hand, the mass of $t'_1$ may be limited from searches for the fourth generation $t'$,
where it is assumed that 
$t'$ is directly produced in $t'\bar t'$ pair and decays exclusively via a specific channel.\footnote{
Here and hereafter, $t'$ denotes a fourth generation up-type quark, while
$t'_i$ does the up-type vector-like quarks of the present model.}


Assuming the exclusive decay $t'\to tZ$, the CMS collaboration obtained a mass bound $m_{t'}>475\GeV$ at 95\% CL with an integrated luminosity of $1.14\invfb$~\cite{Chatrchyan2011svl}, as shown in Fig.~\ref{fig:CMSboundsTZ}.
In the figure we also show $\sigma(pp\to t'_1\bar t'_1)\times [\Br(t'_1\to tZ)]^2$
for each benchmark point.
The total cross section $\sigma(pp\to t'_1\bar t'_1)$
 is calculated with HATHOR~\cite{HATHOR} at the NLO level, using CT10~\cite{PDFCT10} parton distribution function.
It gives constraints only for the points (D) and (E); $m_{t'_1}\simeq 300\GeV$ and
$m_{t'_1}\simeq 265$--$325\GeV$ are excluded at 95\% CL, respectively.
The bound for (E) is obtained because $\Br(t'_1\to tZ)$ is large, while that for (D) is because the $t'_1\to th$ channel is closed and  $\Br(t'_1\to tZ)$ is enhanced  for $m_{t'_1}<298\GeV$.
Thus, the search for $t'\to tZ$ has little sensitivity to Point (D). The bound for (E) will be tightened by accumulating more data.

Two results of the search for $t'$ quarks with $\Br(t'\to bW)=1$ are also presented by the CMS collaboration~\cite{CMSPASEXO11050,CMSPASEXO11051}.
The result is shown in Fig.~\ref{fig:CMSboundsBW}.
The red solid line ``di-lep'' in the figure,  comes from the events in which both of two $W$ bosons decay leptonically ($e\nu_e$ or $\mu\nu_\mu$), where $1.14\invfb$ data is used~\cite{CMSPASEXO11050}.
The blue solid line ``semi-lep'', is extracted from the events with semi-leptonic $W$ decay, i.e. one of the $W$ bosons decays leptonically and the other decays into jets~\cite{CMSPASEXO11051}, using 0.573--0.821$\invfb$ data.
Among the benchmark points, the point (A), where $\Br(t\to bW)=1$, receives a 95\% CL lower bound $m(t'_1)>450\GeV$.

The ATLAS collaboration also presented a result of the search for $t'$ quarks decaying via $t'\bar t'\to qW\bar qW\to q\bar l\nu \bar ql\bar\nu$ with $q=u,d,c,s,b$ using $37\invpb$ data~\cite{ATLAS2011022}.
Because of the looser mass bound due to less integrated luminosity, we do not include this bound in figures.

\begin{figure}[t]
\begin{center}
 \includegraphics[width=0.47\textwidth]{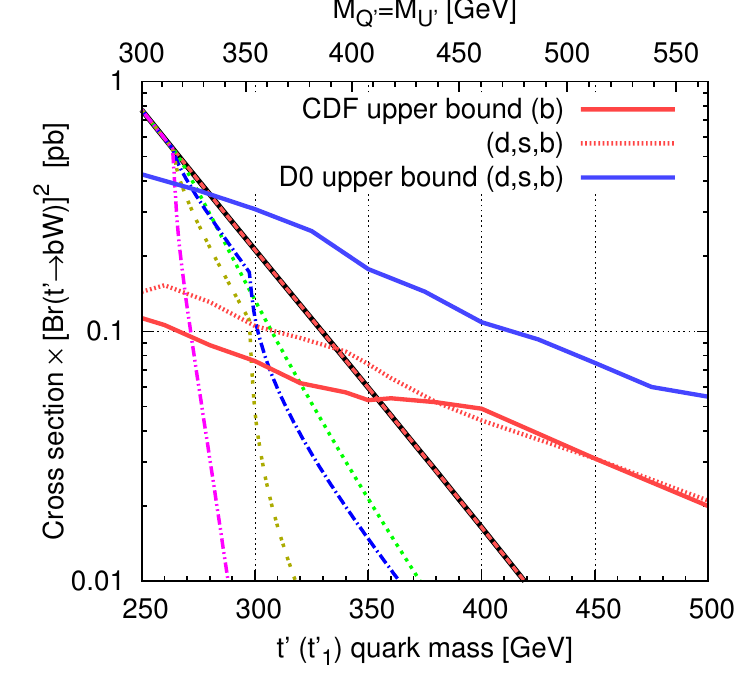}
 \caption{%
The same as Fig.~\ref{fig:CMSboundsBW}, but for the Tevatron experiments. $t'$-quark decay is assumed to be via $t'\to bW$ channel, and the cross section corresponds to $1.96\TeV$ $p\bar p$ collision.
The red-solid, red-dotted, and blue-solid lines denote the upper bounds; see the text for detail.
The other lines are the same as Figs.~\ref{fig:CMSboundsTZ}--\ref{fig:CMSboundsBW}, but for Tevatron cross section.
Note again that the line for (A) sits on that of the total cross section.}
 \label{fig:TEVATRONboundsBW}
\end{center}
\end{figure}

Now let us focus on the Tevatron experiments.
Both the CDF and D0 collaborations published mass bounds on $t'$ quarks decaying via the $t'\to bW$ channel.
As shown as the red-solid line in Fig.~\ref{fig:TEVATRONboundsBW}, the CDF collaboration~\cite{Aaltonen:2011tq} gives a bound $m_{t'}>358\GeV$ at 95\% CL.
They also presented a bound for $t'$ quarks decaying via $t'\to q_dW$ channel, where $q_d$ is a generic down-type quark in the SM (the red-dotted line in the figure).
The blue-solid line shows the bound for $t'\to q_d W$ obtained by the D0 collaboration~\cite{Abazov:2011vy}.

In summary, we would like to emphasize that the Tevatron experiments give mass bounds for all the benchmark points.
This is because $t'_1$ whose mass is less than $m_t+m_Z=264\GeV$ decay exclusively via the channel $t'\to bW$ as is discussed above, and thus the $t'_1$ quark below $264\GeV$ is excluded by the experiments regardless of the mixing parameter, or the branching ratio.
However, as is clear from the results of the LHC experiments especially for Point (D), it is difficult to tighten the mass bound only from the searches for $t'\to tZ$ and $t'\to bW$.
Now we are ready to discuss prospects for further search.

\begin{figure}[t]
\begin{center}
 \includegraphics[width=0.47\textwidth]{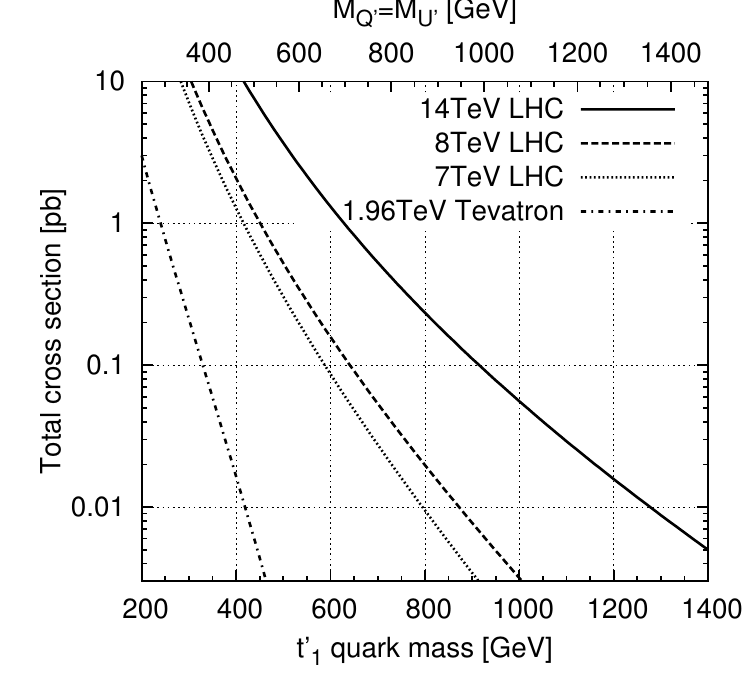}
 \caption{%
The production cross sections of $pp (p\bar p)\to t'_1 \bar t'_1$ as functions of the $t'_1$ quark mass, calculated with HATHOR~\cite{HATHOR} at NLO level.}
 \label{fig:ttbarCS}
\end{center}
\end{figure}
\subsubsection{Prospects of further searches}
First let us review prospects for the search capability at the benchmark points (A)--(E).
Currently, all of the benchmark points receive lower bounds on the mass
from the $t'\to bW$ search by the CDF experiments as shown in Fig.~\ref{fig:TEVATRONboundsBW}.
However, as these bounds owe much to the closure of the other decay channels, further search on this channel has little sensitivity for points (D) and (E).
For the point (E), where $t'_1\to tZ$ is the dominant decay channel, the $t'\to tZ$ search will give a bound soon. On the other hard, the point (D) needs searches for $t'$ quark with $t'\to th$ decay channel, for which no result is published yet.

As the branching ratio depends largely on the ratio among the small mixing parameter, it is of great importance to search for all of the decay modes, especially $t'_1\to th$ channel.
This channel is promising for $t'$ search, because the Higgs boson around $125\GeV$ is expected to decay into $b\bar b$ channel
 and thus three $b$-quarks are expected from one $t'_1$, or more than three in an event.
This characteristic signature would reduce background events and, up to $b$-tagging efficiency, a good signal over background ratio is expected.

Search for the heavier vector-like quarks, $b'$ and $t'_2$ is also well worth doing.
These quarks finally decay into one (or three) $b$ quark(s) and at least one vector boson, so there are at least two $b$ quarks and at least two vector bosons.
Since the event has so many particles, analyses with detector simulation are important for future studies.

The LHC upgrade to $14\TeV$ is expected to provide good sensitivity to the vector-like quarks. As shown in Fig.~\ref{fig:ttbarCS}, the $t'\bar t'$ production cross section would be more than ten times larger than the $7\TeV$ LHC.
This will be delightful especially for the heavier vector-like quark search.

\section{Summary and discussion}
\label{sec:summary}

We have investigated the GMSB models with vector-like matters,
paying particular attention to the Higgs mass, muon $g-2$, and the  LHC discovery prospects of SUSY particles.
In the region where the Higgs boson mass is 124--126 GeV
and the muon $g-2$  is consistent with the experimental 
value at the 1$\sigma$ (2$\sigma$) level, there is an upper bound on the gluino mass,
$m_{\tilde g}\lsim 1.2\TeV$ ($m_{\tilde g}\lsim 1.8\TeV$),
and an upper bound on the extra vector-like quarks, 1.0 (1.8)\,TeV.
The NLSP is either stau or neutralino,
and the LHC prospects of both cases have been studied.
Some parameter regions are already excluded by 7\,TeV LHC, and 
14\,TeV LHC is expected to cover most of the parameter space. 
The LHC search for extra vector-like quarks was also discussed.

In this paper, we have mainly investigated the case where the NLSP is long-lived.
Let us briefly discuss the case where the gravitino is very light, and hence 
the NLSP decays inside the detectors.
In the case of the stau NLSP, in-flight decays of the stau leave kink signatures.
For instance, in the model points LLP1 an LLP2 in Table.~\ref{tab:llp-mass},
the gluino mass is about $1.8-1.9\TeV$ and the stau mass is about $230-280\GeV$.
According to the study in Ref.~\cite{Asai:2011wy}, for such a mass spectrum, 
more than 10 kink events can be observed at the 14\,TeV LHC with an integrated luminosity 10$\invfb$,
for the stau decay length
$c\tau\sim 100$--$4000$ mm. For the stau mass $230-280\GeV$,
this corresponds to the gravitino mass $0.6-6$\,keV.
Since LLP1 and LLP2 are the points with a relatively large gluino mass 
within the parameter region where a $124-126\GeV$ Higgs boson and a muon $g-2$  are simultaneously explained, we expect that most of the parameter space can be covered at 14\,TeV LHC
also in the case of in-flight stau NLSP decay. 
In the case of neutralino NLSP with a very light gravitino, typical LHC signatures will be a 
non-pointing photon~\cite{Kawagoe:2003jv} and/or a neutralino in-flight decay into $Z$-boson~\cite{Meade:2010ji}. Since there is an upper bound on the gluino mass $m_{\tilde{g}}\lsim 1.2\TeV$ (1.8\,TeV) in the case of neutralino NLSP once a $124-126\GeV$ Higgs and the muon $g-2$ constraints at $1\sigma$ ($2\sigma$) are imposed, it is expected that a large part of the parameter space is reached also in this case.

Concerning the vector-like quark, we again emphasize the importance of the search for $t'\to th$, the fourth generation quarks decaying into a $t$ quark and a Higgs boson, especially if the Higgs boson is really discovered around $125\GeV$.
The signature of this decay channel, three $b$ quarks and a $W$ boson, could be observed in the LHC, and thus expected to give a tight bound on the parameters on this model.


The present model is one of the most attractive and phenomenologically viable SUSY models which can explain the Higgs mass 124--126 GeV. Perturbative coupling unification is realized. 
Dangerous flavor/CP violating soft terms are naturally suppressed, 
while the muon $g-2$ can be explained. The model is also cosmologically viable.
The gravitino can be the dominant component of the dark matter with a reheating temperature $T_R \sim 10^8\GEV (m_{3/2} / 1\GEV)^{-1}$~\cite{hep-ph/0012052}.\footnote{A technically important issue in the present setup is that, since the SU(3) gauge coupling remains strong up to the high energy scale, the perturbation in the calculation of the gravitino production becomes troublesome.
(cf.~\cite{hep-ph/0012052}.)
An improvement is necessary to reliably calculate the gravitino abundance in the present model, but it is beyond the scope of this paper.} In such a case, non-thermal leptogenesis~\cite{hep-ph/9906366+X} can explain the cosmological baryon asymmetry.
Furthermore, the notorious inflaton-induced gravitino problem~\cite{arXiv:0706.0986}, which excludes
most of the inflation models in gravity-mediated SUSY breaking models, can also be avoided in the present model.

\section*{Acknowledgments}
NY would like to thank Caltech TH group,
where part of this work has been carried out.
This work was supported by Grand-in-Aid for Scientific research from
the Ministry of Education, Science, Sports, and Culture (MEXT), Japan,
No. 23740172 (M.E.), 
No. 21740164 (K.H.), No. 22244021 (K.H.) and No. 22-7585 (N.Y.).
S.I. is supported by JSPS Grant-in-Aid for JSPS Fellows.
This work was supported by World Premier International Research Center Initiative (WPI Initiative), MEXT, Japan.


\end{document}